\documentclass[
                a4paper,
              ]{tmplts/llncs}

\usepackage{etex}
\usepackage{ifthen} 
\newboolean{blindauthors} 
\setboolean{blindauthors}{false}
\newboolean{extended} 
\setboolean{extended}{false}
\usepackage[utf8]{inputenc} 
\usepackage[english]{babel}
\usepackage{fixltx2e}	\usepackage[protrusion=true, expansion=true]{microtype} \usepackage{ellipsis}	
\usepackage{cite}
\usepackage[numbers, sort&compress, sectionbib]{natbib}

  \usepackage[pdftex]{graphicx}
  \graphicspath{{../figs/}}
  \DeclareGraphicsExtensions{.pdf,.jpeg,.png}

\usepackage[caption=false, font=footnotesize]{subfig}
\usepackage{wrapfig}
\usepackage{sidecap}

\usepackage[cmex10]{amsmath}
\usepackage{amssymb}
\usepackage[cmex10]{mathtools}
\usepackage{amsfonts}
\usepackage{algorithmic}
\usepackage{array}
\usepackage{mdwmath}
\usepackage{mdwtab}

\usepackage{multirow}
\usepackage{booktabs}
\usepackage{paralist}
\usepackage{tabularx}

\usepackage{xspace}
\usepackage{url}
\usepackage{nccfoots}
\usepackage{todonotes}

\usepackage[thmmarks,framed]{ntheorem}

\newlength{\stdJot}
\newlength{\stdParskip}
\newlength{\stdParindent}
\newcommand{\keywords}[1]{
    \par\addvspace\baselineskip
    \noindent\keywordname\enspace
    \ignorespaces
    #1
}

\definecolor{darkgreen}{rgb}{0,0.5,0}
\definecolor{darkblue}{rgb}{0,0,0.5}
\definecolor{darkred}{rgb}{0.5,0,0}

\newcommand{\srefsec}[1]{Sec.~\ref{sec:#1}}    
\newcommand{\sreffig}[1]{Fig.~\ref{fig:#1}}

\newcommand{\eg}{e.\,g.\xspace}
\newcommand{\ie}{i.\,e.\xspace}
\newcommand{\cf}{cf.\xspace}
\newcommand{\wrt}{w.\,r.\,t.\xspace}

\DeclareMathSymbol{\mlq}{\mathord}{operators}{``}
\DeclareMathSymbol{\mrq}{\mathord}{operators}{`'}

\ifthenelse{\boolean{blindauthors}}
{\newcommand{\blinded}[1]{\ldots blinded for review\ldots}}
{\newcommand{\blinded}[1]{#1}}

\ifthenelse{\boolean{blindauthors}}
{\newcommand{\rmvIfBlinded}[1]{\unskip}}
{\newcommand{\rmvIfBlinded}[1]{#1}}

\ifthenelse{\boolean{blindauthors}}
{\newcommand{\chngCiteIfBlinded}[1]{\cite{blindedForReview}}}
{\newcommand{\chngCiteIfBlinded}[1]{#1}}

\ifthenelse{\boolean{blindauthors}}
{}
{}

\newcommand{\authorblock}[1]{\blinded{#1}}

\ifthenelse{\boolean{blindauthors}}
{
\newcommand{\acknowledgementSection}[1]{
\section*{Acknowledgement}
\label{sec:acknowledgement}
\footnotesize{This paper has partially been supported by \ldots blinded for
review \ldots}
}

\newcommand{\acknowledgementFootnote}[1]{
\Footnotetext{}{
\scriptsize{
\footnotesize{This paper has partially been supported by \ldots blinded for
review \ldots}
} } }

\newcommand{\acknowledgementCommand}[1]{
\begin{acknowledgement}
\footnotesize{This paper has partially been supported by \ldots blinded for
review \ldots}
\end{acknowledgement}               
}
}
{\newcommand{\acknowledgementSection}[1]{
\section*{Acknowledgement}
\label{sec:acknowledgement}
{#1}
}

\newcommand{\acknowledgementFootnote}[1]{
\Footnotetext{}{
\scriptsize{#1} 
} }

\newcommand{\acknowledgementCommand}[1]{
\begin{acknowledgement}
\footnotesize{#1}
\end{acknowledgement}               
}
}

\begin{document}
\mainmatter

\title{
Structural Contracts
}

\subtitle{
Contracts for Type Construction \& Dependent Types 
\\
to Ensure Consistency of Extra-Functional Reasoning
}

\toctitle{
Structural Contracts --
Contracts for Type Construction \& Dependent Types 
to Ensure Consistency of Extra-Functional Reasoning
}

\author{ \authorblock{
    Gregor Nitsche } }

\tocauthor{ \authorblock{
    Gregor Nitsche  
} }

\institute{ \authorblock{
\vspace{-\baselineskip}
    Carl von Ossietzky University Oldenburg, Oldenburg 26121, DE
    \\
    \email{\url{
    gregor.nitsche
    @offis.de}}
} }

\maketitle

\acknowledgementFootnote{\noindent
This research is funded by the German Research Foundation (DFG) in the Research
Training Group SCARE – System Correctness under Adverse Conditions (DFG-GRK
1765).

}

\begin{abstract}
\vspace{-\baselineskip}

Targeting to use contract-based design for the specification and refinement of
extra-functional properties, this research abstract suggests to use type
constraints and dependent types to ensure correct and consistent top-down
decomposition of contracts with respect to a specifiable type constructor. For
this, we summarize the composition problem and give a short draft of our
approach, called \emph{structural contracts}.

\end{abstract}

\subsubsection{Motivation.}
\label{sec:introduction}

\vspace{-0.7\baselineskip}

Design hierarchy and correct hierarchical reasoning are of great importance to
manage the complexity of today's system designs. By refining the system
requirements -- into a subsystem architecture and requirements of its parts --
the complexity can be split up and the probability of design faults can be
reduced.  For this purpose, \emph{Contract Based Design (CBD)} \cite{BCN+18} is
a formalism, which enables to specify and to formally verify such refinements.

In our approach we wish to reason about \emph{extra-functional properties
(EFPs)}, by which we mean the formal specification and verification of
properties and detection of faults which depend on multiple physical domains
(\emph{quantities}), as \eg power consumption.
To ensure safety, today's distributed and reuse-oriented development processes
must be able to reliably detect and prevent such faults when integrating
electronic sensors and controllers within a safety-critical system. 

Our goal is to use CBD for reasoning about these EFPs. For this, a specification
and verification of domain-specific composition functions becomes necessary. We
believe, this is not sufficiently considered in CBD, yet. Accordingly, this
abstract gives an outline of this problem as well as of our approach of
\emph{structural contracts} (SCs), oriented on \emph{dependent types} from
\emph{constructive type theory}. \cite{wwwTypeTheory}

\vspace{-\baselineskip}

\subsubsection*{Outline of the Preliminaries.}
\label{sec:preliminaries}

\begin{wrapfigure}[8]{r}{0.35\columnwidth}
\centering
\vspace{-13mm}
\includegraphics[width=0.35\columnwidth]{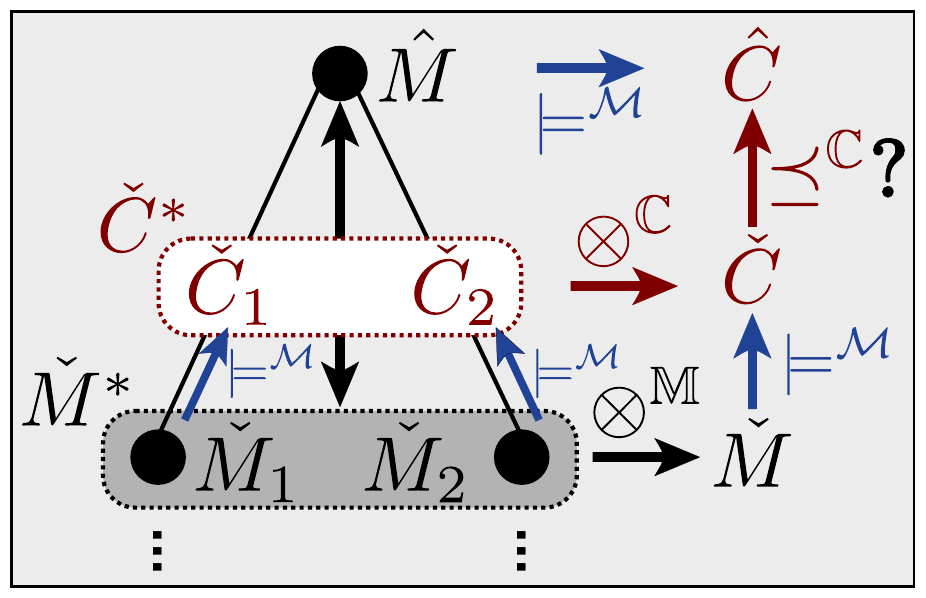}
\vspace{-8mm}
\caption{
\footnotesize
Hierarchical Refinement with Contracts.
}    
\label{fig:hierarchySchema}
\end{wrapfigure}

The overall goal and concept of the CBD approach \cite{BCN+18} is to specify and
refine the requirements of a system $\hat{M}$ based on a contract $\hat{C}$ into
$n \in \mathbb{N}$ contract-based specifications $\check{C}_{k \in K_n}$ of a
set of $k \in K_n$-indexed subsystems $\check{M}^* \Coloneqq \{\check{M}_1,
\check{M}_2, \ldots, \check{M}_n\}$ (where $K_n \Coloneqq \{1, \ldots, n\}
\subseteq \mathbb{N}$), denoting its \emph{components}. 
\sreffig{hierarchySchema} gives a draft of this concept, sketching the
refinement of a system $\hat{M}$ with the contract based specification $\hat{C}$
and its refinement into subsystems $\check{M}_k \in \check{M}^*$ with contracts
$\check{C}_k$.
Composing the components' specifications $\check{C}_k$ to the composed
specification $\check{C} \Coloneqq \otimes^\mathbb{C}_k \check{C}_k$ of the
composed system $\check{M} \Coloneqq \otimes^\mathbb{M}_k \check{M}_k$ (with
$\otimes^\mathbb{M}_k$ denoting k-ary composition over the set $\mathbb{M}$ of
components and $\otimes^\mathbb{C}_k$ denoting k-ary composition over the set
$\mathbb{C}$ of contracts) the refinement relation $\check{C} \preceq^\mathbb{C}
\hat{C}$ can be verified.

For this, a \emph{contract} $C \Coloneqq (A,G)$ formally describes a requirement
by separated assertions $A$ and $G$, denoting the \emph{assumptions} $A$ a
component expects from its embedding environment, plus the corresponding
\emph{guarantees} $G$, which are provided for the case that the assumptions hold
(formally ${A \rightarrow G}$). 
Semantically, the contracts are interpreted into $\mathbb{M}$ such that
$[\![\cdot]\!]^\mathbb{C} : \mathbb{C} \rightarrow \mathfrak{P}^2_\mathbb{M}$ 
with
$[\![C]\!] \Coloneqq (\mathcal{E}, \mathcal{M})$
where $\mathfrak{P}_\mathbb{M}$ denotes the powerset over $\mathbb{M}$ and
$\mathcal{E}, \mathcal{M} \subseteq \mathbb{M}$ denotes the contracts' sets of
compatible \emph{environments} and consistent \emph{implementations},
respectively. 
Furthermore, two satisfaction relations $\models^\mathcal{E},
\models^\mathcal{M} \subseteq \mathbb{M} \times \mathbb{C}$ are defined such
that 
${(E \models^\mathcal{E} C)} \Leftrightarrow {(E \in \mathcal{E})}$ 
and
${(M \models^\mathcal{M} C)} \Leftrightarrow {(M \in \mathcal{M})}$. 
Based on that, to compositionally reason about contracts, the composition
operation $\otimes^\mathbb{C}_{k \in K_n}: \mathbb{C}^n \rightarrow \mathbb{C}$
is defined for $n$ contracts $\check{C}_k$, such that
$\otimes^\mathbb{C}_{k \in K_n} \check{C}_k \Coloneqq$\\[1mm]
\makebox[\textwidth][c]{\begin{scriptsize}
$
\textit{min} \left\lbrace 
\!{\check{C} \!\in\! \mathbb{C}}\!
:\hspace{-1.1mm}
\left[ \begin{array}{l}
\forall \hat{E} \!\in\! \mathbb{M},\\ \forall {\check{M}_{k \in K_n}} \!\in\! \mathbb{M}
\end{array} \right]
\hspace{-1.1mm}:\hspace{-1.1mm}
\left[ \begin{array}{l}
{(\hat{E} \models^{\mathcal{E}}\! \check{C})} 
{\land}\\\bigwedge_{k \in K_n}\!
{(\check{M}_k \models^{\mathcal{M}}\! \check{C}_k)} 
\end{array} \right]
\hspace{-1.1mm}\rightarrow\hspace{-1.1mm}
\left[ \begin{array}{l}
{\bigwedge_{k \in K_n}\! ( (\hat{E} \otimes^\mathbb{M}\! ( \otimes^\mathbb{M}_{(j
\neq k) \in K_n} \check{M}_j ) ) \models^{\mathcal{E}}\! \check{C}_k )}
\\\land\,
{(\otimes^\mathbb{M}_{k \in K_n}\! \check{M}_k \models^{\mathcal{M}}\! \check{C})} 
\end{array} \right]
\right \rbrace
$
\end{scriptsize}}

\noindent 
Please note that the composition operation $\otimes^\mathbb{M}: \mathbb{M}^2
\rightarrow \mathbb{M}$ (resp. $\otimes^\mathbb{M}_k: \mathbb{M}^n \rightarrow
\mathbb{M}$) over components is described as a part of our problem statement
(\cf \srefsec{problems}).

With that, the three main verification rules of CBD are: checking
\emph{compatibility} ($\mathcal{E} \neq \varnothing$ ?) and \emph{consistency}
($\mathcal{M} \neq \varnothing$ ?) of all contracts, and checking
\emph{refinement} 
($
(\check{C} \preceq^\mathbb{C} \hat{C}) 
\Leftrightarrow
(
(\hat{\mathcal{E}} \subseteq \check{\mathcal{E}}) 
\land 
(\hat{\mathcal{M}} \supseteq \check{\mathcal{M}})
)
$)
between the system specification $\hat{C}$ and the composed specification
$\check{C} = \otimes^\mathbb{C}_k \check{C}_k$  of the composed subsystems
$\check{M}_k \in \check{M}^*$.

\subsubsection{Problem Description.}
\label{sec:problems}

\vspace{-1.3\baselineskip}

\begin{wrapfigure}[9]{r}{0.47\columnwidth}
\centering
\vspace{-2\baselineskip}
\includegraphics[width=0.48\columnwidth]{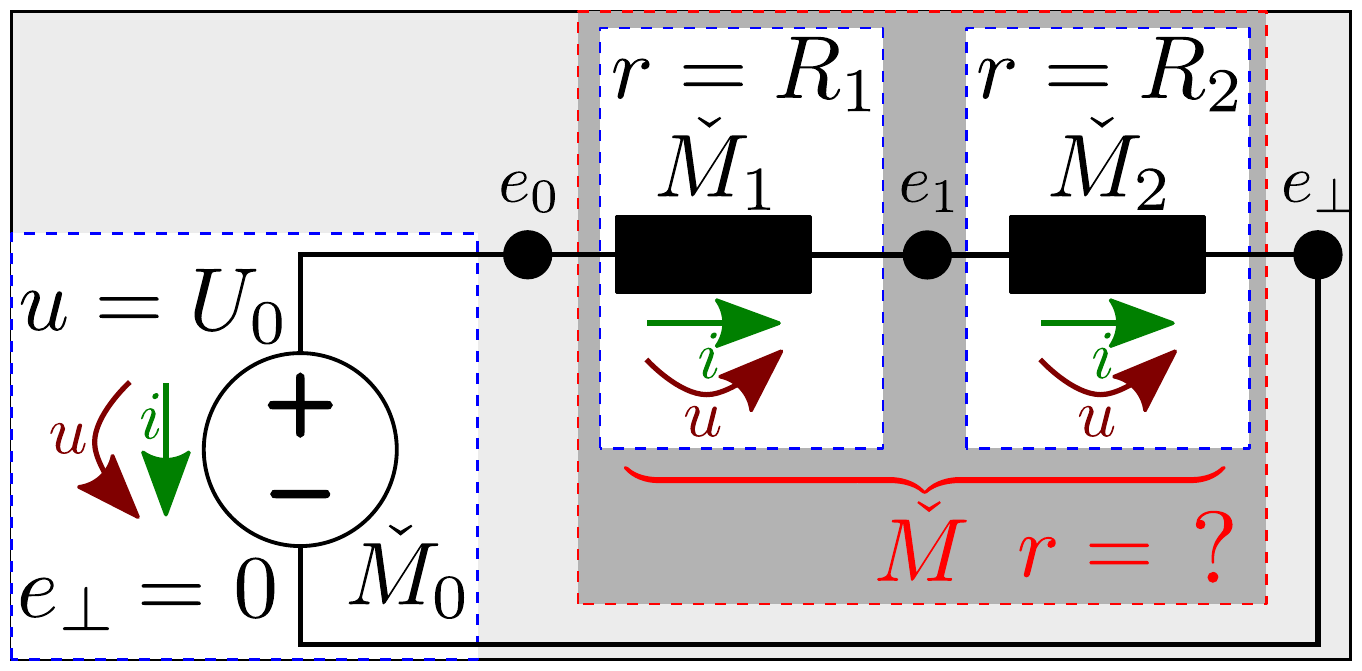}
\vspace{-7.4mm}
\caption{
\footnotesize
Example: Composing resistors.
}    
\label{fig:cmpExample}
\end{wrapfigure}

We wish to introduce our problem based on an example from electronics (\cf
\sreffig{cmpExample}) where we compose two resistors $\check{M}_1$,
$\check{M}_2$ in series, resulting in a composed resistor $\check{M}$.
For this composition, we examine how the equivalent electrical resistance
$\check{M}.r$ results from $\check{M}_1.r$ and $\check{M}_2.r$.
Considering $\mathbb{M}$ to reflect components from simple electrodynamic
circuit theory, those follow \emph{Ohm's Law} ($R=u/i$) for resistors,
constraining the voltage $u$ across and the current $i$ through any resistor.
Furthermore, each node $e \!\in\! \{e_0, e_1, e_\bot\}$ satisfies
\emph{Kirchhoffs Current Law} ($\sum_{k \in K_e} i_k\!=\!0$), where $K_e
\!\subseteq\! \{0,1,2\}$ defines the subsets $M_e \Coloneqq \{\check{M}_k : k
\!\in\! K_e\}$ of the components that are connected by $e$. 
Finally, due to \emph{Kirchhoffs Voltage Law} and the \emph{Law of
Superposition}, the \emph{branches} $b \in \{(e_j, e_k) : j,k \in \{0, 1,
\bot\}\}$ satisfy ${u_{e_j} \!-\! u_{e_k}} = u_m$ for all branch components $M_m
\!\in\! {K_{e_j} \!\cap\! K_{e_k}}$.
Abbreviating (sub-)component properties $\check{M}.p$ with $\check{p}$ and
$\check{M}_k.p$ with $\check{p}_k$   
(\ie ${\forall p \in \{r, u, i\}} : {\forall k \!\in\! \{0, 1, 2\}} :
{\check{M}_k.p \equiv \check{p}_k}$), 
we can now show: $(\check{r} = \check{u}/\check{i}) \Leftrightarrow (\check{r} =
\check{r}_1 + \check{r}_2)$ since $\forall k \in \{1,2\}: \check{r}_k =
{\check{u}_k/\check{i}_k}$, ${\check{i}_1 = \check{i}_2}$ and $\check{u} =
{\check{u}_1 + \check{u}_2}$. 
Similarly, we can show: $\check{\mathcal{P}} = {\check{\mathcal{P}}_1 +
\check{\mathcal{P}}_2}$ for the power dissipation (defined as $\mathcal{P} = u
\cdot i$) or $\check{r} = (1/\check{r}_1 + 1/\check{r}_2)^{-1}$ for parallel
composition of resistors. 

As we explained in the preliminaries, contracts are interpreted by
tuples $(\mathcal{E}, \mathcal{M})$ of subsets of $\mathbb{M}$. 
Thus, two semantically different, valid assertions $G_1 \!\neq\! G_2$ must be
interpreted to two different subsets $\mathcal{M}_1 \!\neq\! \mathcal{M}_2
\!\subset\! \mathbb{M}$.
Vice versa: to distinguish two different (sets of) components $\mathcal{M}_1
\!\neq\! \mathcal{M}_2 \!\subset\! \mathbb{M}$ by contracts $(A,G)$, the syntax and the
semantics of $A$ and $G$ must be able to describe and reflect the specific
distinction.
Thus, if a difference in the composition operation is relevant for
the validity of composing contracts (and checking their refinement), the
assumptions and guarantees must be able to catch this. 
For our example, if we want a valid serial decomposition of $\check{r}$, it
would not be sufficient to specify only a value (\eg $\check{r} \!=\! 3$) for the
resistance, since -- \eg having subcomponents $\check{M}_1, \check{M}_2$ with
$\check{r}_1 \!=\! 1$ and $\check{r}_2 \!=\! 2$ -- there are several composition
operations which erroneously would satisfy $\check{r} \!=\! 3$, \eg $\check{r}
= 3 \check{r}_1 \!+\! 0 \check{r}_2$
While those composition operations have same \emph{type signatures}
$\mathbb{M} \!\times\! \mathbb{M} \!\rightarrow\! \mathbb{M}$, mapping $n\!=\!2$
resistors $\check{M}_k \!\in\! \mathbb{M}$ to a resistor $\check{M} \!\in\!
\mathbb{M}$ their \emph{term signatures} (\eg $3 \check{r}_1 \!+\! 0 \check{r}_2
\not\equiv
\check{r}_1 + \check{r}_2$) differ.

However, in the definition of the composition operation $\otimes^\mathbb{C}_{k
\in K_n}$ for contracts (\cf preliminaries), CBD requires a contract
theory to provide a fixed composition operation $\otimes^\mathbb{M}_{k \in
K_n}$, whose type and term signature is valid for all $\check{M}^* \in
\mathbb{M}^n$.
This limits the applicability of contract based design to components, which do
not need to distinguish their properties by means of the composition term. 
As a consequence, a contract theory can support correct composition (and thus
refinement checking) for only a limited number of $s \in \mathbb{N}$ predefined
component subsets $\mathbb{M}_{k \in K_s} \subseteq \mathbb{M}$ with
$\bigcap_{\forall k \in K_s} \mathbb{M}_k = \varnothing$ and $\bigcup_{\forall k
\in K_s} \mathbb{M}_k = \mathbb{M}$.

\vspace{-1.2\baselineskip}

\subsubsection*{Related Work.}
\label{sec:related_work}

As we first identified the difficulties with properly defining, composing and
refining EFPs in terms of contracts, we outlined a problem sketch in
\cite{NGG+15} and suggested an early approach, based on extending the refining
set $\check{M}^*$ by another subcomponent, having assumptions and guarantees
about the interconnection relations between the other components' variables. 
Most closely related, \cite{PS17} agrees with our idea, but by distinguishing
different contract types and specifying assertion and validity rules (\ie
functions and predicates) between the variables of that contract types. In
contrast to SCs, these rules and types are defined as sets at the level of
\emph{platforms}, limiting the set of correctly refineable property
specifications according to only those rules and types of that platform.
Being based on the concept of type constructors and product types, we belief our
SCs to be a more generic approach that is able to express required type and
composition constraints also between different platforms and to support their
bidirectional propagation across multiple levels of hierarchical
composition.
Furthermore, \emph{vertical contracts} \cite{NS18} propose an \emph{architecture
mapping} relation to relate \emph{aggregation} functions between the two
hierarchy levels of a refinement. However, their \emph{aggregation} function
captures only parallel and conjunctive composition $\otimes^\mathbb{C}_k$
and $\land^\mathbb{C}_k$ of contracts.  
Further related work investigates \eg: (meta)-model-based composition rules for
composing EFPs via contracts \cite{SSC+16}; consistency reasoning for contracts
based on onthologies \cite{VDD+16}; or \emph{Galois Connections} for the
compositional abstraction of contracts \cite{BCN+18}.

\vspace{-0.7\baselineskip}

\subsubsection{Our Approach of Structural Contracts.}
\label{sec:structural_contracts}

\vspace{-0.7\baselineskip}

To allow CBD to correctly reason about an arbitrary number of user-defined
(extra-functional) properties, our goal is to extend CBD by a specification and
verification mechanism -- called \emph{structural contracts} (SCs) -- that
constrains the accepted term and type signatures of component construction and
composition, which means:
a) allowing to control the partitioning of $\mathbb{M}$ into a set $\mathbb{M}_S
\Coloneqq \bigcup_{\forall k \in K_s} \{\mathbb{M}_k\}$ of $s$ hierarchically
ordered, (extra-functional) subdomains $\mathbb{M}_{k}$, and b) allowing
to specify the composition operations $\otimes^{\mathbb{M}_S}_{k \in K_n}$,
whose type signatures are defined via the product type $\Pi_{\forall k \in K_n,
\forall \mathbb{M}_k \in \mathbb{M}_S} \mathbb{M}_{k} \rightarrow
\check{\mathbb{M}} \in \mathbb{M}_S$. 
Thus, following the \emph{Curry-Howard-Isomorphism} of `propositions-as-types'
\cite{wwwTypeTheory} from \emph{constructive type theory}, a similarly
comprehensive `structural' notion of contracts (`contracts-as-types') requires
several extensions \wrt \emph{polymorphic} and \emph{dependent} types:
\vspace{-0.3\baselineskip}
\begin{enumerate}
\item language extension of contracts, to allow for a specification of
    polymorphic and dependent type requirements (\ie the syntax of our SCs);
\item semantical extension from a set $\mathbb{M}$ of components to a
    hierarchically ordered, extensible set of component-types $\mathbb{M}_{k}
    \!\in\! \mathbb{M}_S$, based on type- \& term-dependent type-constructors
        (\ie the type-theoretic semantics of our SCs);
                            \item implement and integrate algorithms for the corresponding type checking;
\end{enumerate}
\vspace{-0.3\baselineskip}
For brevity, we here focus only on giving an outline of point 2:
To enable polymorphic and dependent product types for contracts, we must be able
to overload the composition operation $\otimes^{\mathbb{M}_S}_k$ with an
appropriate, user definable term signature for each type signature of the
product type $\Pi_{\forall k \in K_n, \forall \mathbb{M}_k \in \mathbb{M}_S}
\mathbb{M}_k$.

For our example, this would mean the declaration of the type `resistance' as
$\mathbb{M}_r: \bot \rightarrow \mathbb{M}$, which can be used for the
generation of the type 'resistor' $\mathbb{M}_{M_r}: \mathbb{M}_r \rightarrow
\mathbb{M}$, \ie components with a property $r$ of type $\mathbb{M}_r$ denoted
by $r : \mathbb{M}_{r}$.
With that, the type signature of our composition operation can become
$\otimes^{\mathbb{M}_r} : (\mathbb{M}_r \times \mathbb{M}_r) \rightarrow
\mathbb{M}_r$ or generally $\otimes^{\mathbb{M}_r} : (\Pi_{k \in \mathbb{N}}
\mathbb{M}_r^k) \rightarrow \mathbb{M}_r$.
Furthermore, to define new (sub-)types within the type hierarchy of
$\mathbb{M}_S$, we suggest to have type constructors $\mathbb{M}_k: \bot
\rightarrow \mathbb{M}_S \cup \{\mathbb{M}_S\}$, meaning the declaration of the
type $\mathbb{M}_k$ as a base type in $\mathbb{M}_S$ or as a subtype of some
other $\mathbb{M}_{k'} \in \mathbb{M}_S$.

\vspace{-\baselineskip}

\subsubsection{Status and Future Work.}
\label{sec:conclusion}

Investigating CBD for EFPs, we identified difficulties \wrt the ability to
specify user-defined property-domains and composition operations. As a promising
approach, we now combine CBD with dependent types and constraints for the type
constructors. After finishing the theoretic fundament and a first
implementation, we plan to evaluate our SCs for different EFPs of embedded,
electrical systems, like timing, power consumption and heating.

\vspace{-\baselineskip}

\bibliographystyle{tmplts/splncs04}
\renewcommand\bibname{
{\vspace{-0.5\baselineskip}
\normalsize
{\kern-0.4em References}
}
}
\bibliography{paper_main.bbl}

\ifthenelse{\boolean{extended}}
{
\newpage
\begin{appendix}
\input{annex/annex1}
\end{appendix}
}{}

\end{document}